# The misuse of the nonlinear field normalization method: Nonlinear field normalization citation counts at the paper level cannot be added or averaged

**Xing Wang[a],[1]**

[a]School of Information, Shanxi University of Finance & Economics, 030006 Taiyuan, China

[1]Corresponding author. Email: wangxing@sjtu.edu.cn

**Abstract** There is a very important problem that has not attracted sufficient attention in academia, i.e., nonlinear field normalization citation counts at the paper level obtained using nonlinear field normalization methods cannot be added or averaged. Unfortunately, there are many cases adding or averaging the nonlinear normalized citation counts of individual papers that can be found in the academic literature, indicating that nonlinear field normalization methods have long been misused in academia. In this paper, we performed the following two research works. First, we analyzed why the nonlinear normalized citation counts of individual papers cannot be added or averaged from the perspective of theoretical analysis in mathematics: we provide mathematical proofs for the crucial steps of the analysis. Second, we systematically classified the existing main field normalization methods into linear and nonlinear field normalization methods. The above two research works provide a theoretical basis for the proper use of field normalization methods in the future, avoiding the continued misuse of nonlinear data. Furthermore, because our mathematical proof is applicable to all nonlinear data in the entire real number domain, our research works are also meaningful for the whole field of data and information science.

**Keywords** field normalization method; nonlinear normalization method; nonlinear transformation; equidistant transformation

## Introduction

Citation count is an important and commonly used quantitative indicator as a proxy for the research impact of papers in the scientific community, involving research impact evaluation at aggregation levels (e.g., at the level of individual scholars, research groups, institutions, journals and countries). The outcome of such an evaluation often plays an important role in deciding the granting of research funds, hiring of scholars and even the fate of scientific institutions. Therefore, research impact evaluation based on citation analysis should be performed in the most precise and unbiased manner (Radicchi et al., 2008).

However, one potential problem of research impact evaluation based on citation analysis is that the raw citation counts of papers from different fields cannot be compared directly due to inherent differences in citation practices among fields. Such comparison of raw citation counts is biased and may hamper a fair evaluation of research impact. For example, the citation count of a paper in mathematics is typically lower than that of a paper in molecular biology.

Given this situation, it is necessary to use field normalization to minimize or eliminate field differences to ensure a fair comparison of citation impact across fields[1]. A field normalization method is a mathematical transformation that attempts to eliminate field differences in citation counts. After normalization, the normalized citation counts should be independent of field so that the comparison of citation impact across fields can be naturally implemented through the comparison of normalized citation counts.

To date, various field normalization methods have been proposed with the development of informetrics. These field normalization methods can be classified into linear and nonlinear field normalization methods based on whether the normalization procedure is a linear transformation (Zhang et al., 2015). For instance, one commonly used method in research evaluation practice, the mean-based method, which divides the raw citation count of one paper by the average citation counts of all papers in the same field, is a typical linear field normalization method, while another very popular normalization method, the percentile rank method, which maps the raw citation counts of papers to their relative positions in the citation distribution, is a typical nonlinear field normalization method.

There is a very important problem that has not attracted sufficient attention in academic and evaluation practice circles, i.e., the nonlinear field normalization citation counts at the paper level obtained using nonlinear field normalization methods cannot be added or averaged. Actual research evaluation is often conducted at aggregation levels (e.g., at the level of institutions, journals and countries), which requires adding or averaging the normalized citation counts of individual papers to measure the normalized citation performance at the aggregation level. However, many cases of adding or averaging the nonlinear normalized citation counts of individual papers can be found in the academic literature or evaluation practice circles (e.g., Bornmann, 2020; Bornmann & Williams, 2020; Lundberg, 2007; Maflahi & Thelwall, 2021; Schniedermann, 2021; Thelwall, 2017, 2020; Thelwall & Maflahi, 2020; Wu et al., 2020), indicating that nonlinear field normalization methods have long been misused. We analyze why the nonlinear field normalization citation counts of individual papers cannot be added or averaged in the second section, which is the focus of this paper.

In addition, we comb and summarize existing popular field normalization methods according to whether their normalization procedures are linear transformations in the third section and thus classify these normalization methods into linear and nonlinear field normalization methods. Finally, we offer concluding remarks in the "Conclusions and discussions" section and discuss a dilemma regarding the use of field normalization methods in the Supporting Information, which involves the test method of the normalization effect and the ideal goal of field normalization.

## Why the nonlinear field normalization citation counts of individual papers cannot be added or averaged

### *What is a linear field normalization method and what is a nonlinear field normalization method*

We first define linear and nonlinear field normalization methods.

In the same field, if there is a mapping relationship such as $y = kx + b$ ($k$ is a nonzero constant and $b$ is a constant) between $x$, the raw citation count of a paper, and $y$, the normalized citation count of that paper, then we say that $y$ is the linear transformation of $x$, and this transformation is said to be a linear field normalization method. Otherwise, we say that $y$ is the nonlinear transformation of $x$, and the corresponding transformation procedure is said to be a nonlinear field normalization method.

For example, the popular mean-based method (see equation (1)) and z score method (see equation (2)) are typical linear normalization methods:

$$y = \frac{x}{m} = \frac{1}{m} \cdot x \quad , \tag{1}$$

$$y = \frac{x - m}{sd} = \frac{1}{sd} \cdot x - \frac{m}{sd} \quad . \tag{2}$$

In equation (1), $x$ and $y$ are the raw and normalized citation counts of papers, respectively, and $m$ is the average citation count of papers with the same subject, publication year, and document type as the given paper. In fact, the mean-based method is a specific case of the general linear method $y = kx + b$, where the scaling term $k =$ $1/m$ and the intercept term $b = 0$. Similarly, the meanings of $x$, $y$, and $m$ in equation (2) are the same as those in equation (1), while $sd$ is the standard deviation of the citation counts of papers with the same subject, publication year, and document type as the given paper; the z score method is also a specific case of the general linear method $y = kx + b$, where the scaling term $k = 1/sd$ and the intercept term $b = -m/sd$.

The popular percentile rank method is a typical nonlinear field normalization method. The percentile rank score $y$ of one paper's raw citation count $x$ is defined as the citation counts of $y\%$ of the papers with the same subject, publication year, and document type as the given paper are (at or) below the raw citation count $x$ of the given paper[2]. For instance, let us assume that a given paper has a raw citation count of 30. If 80% of the papers with the same subject, publication year, and document type as the given paper are (at or) below the citation count of 30, then the percentile rank score of the given paper is 80. The mapping from the raw citation count $x$ to the percentile rank score $y$ under the percentile rank method usually cannot form a linear relationship, such as $y = kx + b$; thus, the percentile rank method is a nonlinear field normalization method.

### *"Equal-interval" measurement: the criterion for judging whether the normalized citation counts of individual papers can be added*

One point that people sometimes easily overlook is that addition requires equal-interval measurement, that is, each measuring unit of the data must be equidistant. For example, the distance (i.e., difference) between a test score of 1 point and 2 points is the same as the distance between 2 and 3 points. The equidistance of the measuring unit of data is a prerequisite for the data to be added. If we omit this important prerequisite, the wrong results will be obtained. Let us consider the following fictitious example.

There are two boxes in Figure 1, Box 1 and Box 2. We can use the lengths of people's hands as the measuring unit to measure the lengths of these two boxes. There are two measuring methods. The first method is to use the hands of different people to form a ruler (i.e., Ruler A in Figure 1) to measure the lengths of the two boxes. Because different people have hands of different lengths, the measuring unit here (i.e., the length of each hand) is not equidistant. Under this method of measurement, Box 1 is 3 "hands" length and Box 2 is 9 "hands" length, which will lead us to draw the conclusion that the sum of the lengths of three of Box 1 is equal to the length of Box 2 since "3 hands length+ 3 hands length + 3 hands length = 9 hands length".

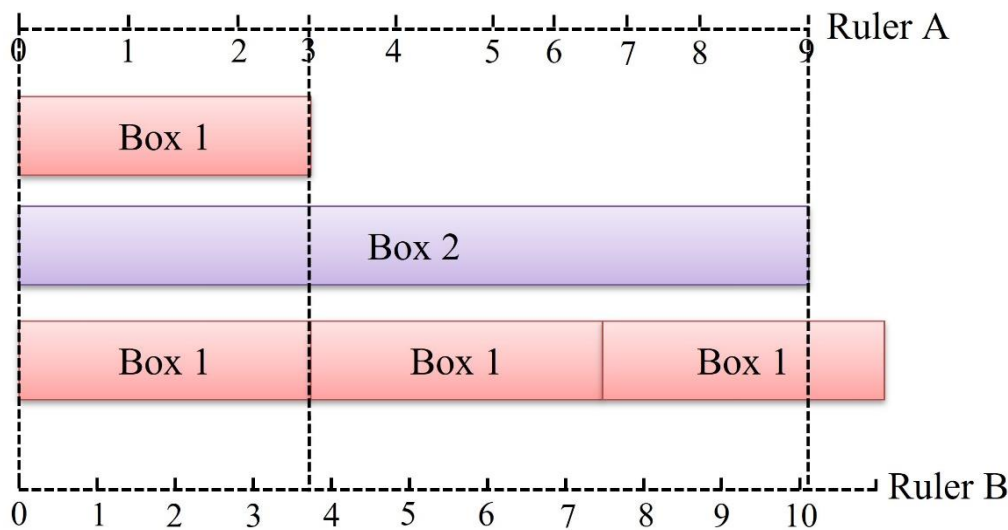

**Figure 1** A fictitious sample to explain why addition requires equal-interval measurement
*Note*: The length of the dashed line "------" between each number denotes the length of people's hands, i.e., the measuring unit.

However, as shown in Figure 1, in reality, the sum of the lengths of three Box 1s is greater than the length of Box 2. Obviously, the conclusion drawn from the first method of measurement is wrong. Unlike the first method of measurement, the second method of measurement uses one hand of the same person to form a ruler (i.e., Ruler B in Figure 1) to measure the lengths of the two boxes. In this way, the length of the hand (i.e., the measuring unit) is fixed; that is, the measuring unit is equidistant. Clearly, using the second scientific method of measurement, we can obtain the correct conclusion about the measurement, which coincides with the real situation.

Thus far, it is easy to understand from the above example why addition requires equal-interval measurement. When the measuring unit of data is not equidistant, the data cannot be added; otherwise, we would obtain the wrong conclusion.

Similarly, each citation count of the raw citation counts of individual papers is equidistant in the same field (in fact, each citation count here is equivalent to the measuring unit in the above examples—the length of hands); thus, raw citation counts in the same field can be added or averaged. However, if we want to add or average the normalized citation counts of individual papers after implementing field normalization, we must also ensure that each measuring unit of these normalized citation counts in the same field is equidistant, which implies that the field normalization must be a linear transformation; we will explain the reason in the next section.

### *The nonlinear normalized citation counts are no longer equidistant*

The normalization procedure of the linear field normalization method is an equidistant transformation that does not change the equidistant nature of each citation count (in fact, the linear normalization transformation and the equidistant transformation are mutually sufficient and necessary conditions; we have proved this theorem in Appendix A). Here, we used a fictitious dataset including 52 papers with various citation counts in a field to demonstrate this point in an easy-to-understand manner (see Table 1).

**Table 1** The dataset of papers for calculating normalized citation counts based on the mean-based method and percentile rank method (52 papers)

| Citation count | Number of papers | $k = 1/m$ | $m$-score | Rank $i$ | Percentile rank score |
|---|---|---|---|---|---|
| 0 | 9 | 0.07 | 0.00 | 9 | 17.31 |
| 1 | 8 | 0.07 | 0.07 | 17 | 32.69 |
| 2 | 6 | 0.07 | 0.14 | 23 | 44.23 |
| 3 | 4 | 0.07 | 0.21 | 27 | 51.92 |
| 4 | 4 | 0.07 | 0.28 | 31 | 59.62 |
| 5 | 2 | 0.07 | 0.35 | 33 | 63.46 |
| 6 | 2 | 0.07 | 0.42 | 35 | 67.31 |
| 7 | 1 | 0.07 | 0.49 | 36 | 69.23 |
| 8 | 1 | 0.07 | 0.56 | 37 | 71.15 |
| 9 | 1 | 0.07 | 0.63 | 38 | 73.08 |
| 10 | 2 | 0.07 | 0.70 | 40 | 76.92 |
| 12 | 1 | 0.07 | 0.84 | 41 | 78.85 |
| 15 | 1 | 0.07 | 1.05 | 42 | 80.77 |
| 16 | 1 | 0.07 | 1.12 | 43 | 82.69 |
| 20 | 2 | 0.07 | 1.40 | 45 | 86.54 |
| 25 | 1 | 0.07 | 1.74 | 46 | 88.46 |
| 38 | 1 | 0.07 | 2.65 | 47 | 90.38 |
| 42 | 1 | 0.07 | 2.93 | 48 | 92.31 |
| 43 | 1 | 0.07 | 3.00 | 49 | 94.23 |
| 80 | 1 | 0.07 | 5.58 | 50 | 96.15 |
| 120 | 1 | 0.07 | 8.38 | 51 | 98.08 |
| 200 | 1 | 0.07 | 13.96 | 52 | 100.00 |

*Note*: $k$ refers to the scaling term in the general linear method $y = kx + b$; $m$ denotes the mean value of the raw citation counts of 52 papers; $m$-score refers to the normalized citation counts obtained using the mean-based method; $i$ denotes the rank of the raw citation counts; the percentile rank score refers to the normalized citation counts obtained using the percentile rank method[3].

In this fictitious example, we used the mean-based method, which is a typical linear field normalization method, to normalize the raw citation counts of 52 papers in the field and obtained the normalized citation counts (i.e., $m$-scores) of these papers (see the fourth column in Table 1). Before implementing the mean-based method, each citation count (i.e., the measuring unit) of raw citation counts is equidistant. After implementing the mean-based method, each measuring unit of normalized citation counts is still equidistant: it is just that each citation count of the raw citation counts is scaled by $k = 1/m$, so the distance between each measuring unit of the normalized citation counts via this scaling all become $1/m$, which is 0.07 in this example (see Figure 2).

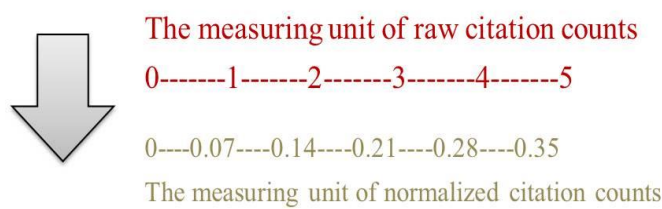

**Figure 2** Schematic diagram of equidistant transformation from raw citation counts to normalized citation counts

Other linear field normalization methods have a similar transformation principle to the mean-based method above: each measuring unit of normalized citation counts via the linear transformation is still equidistant, and this distance is equal to the scaling term $k$ (see Appendix A). Therefore, the linear normalized citation counts of individual papers can be added or averaged.

However, the normalization procedure of the nonlinear field normalization method is not an equidistant transformation; therefore, the nonlinear normalized citation counts are no longer equidistant and thus cannot be added or averaged. Here, we used another example with the same fictitious dataset above to demonstrate this point.

In Table 1, we used the percentile rank method, which is a typical nonlinear field normalization method, to normalize the raw citation counts of the 52 papers and obtained the normalized citation counts (i.e., percentile rank scores) of these papers (see the sixth column in Table 1). Before implementing the percentile rank method, each citation count (i.e., the measuring unit) of raw citation counts is equidistant. After implementing the percentile rank method, each measuring unit of the normalized citation counts is no longer equidistant. For example, the distances between 0 citations and 1 citations, 3 citations and 4 citations, and 42 citations and 43 citations for the raw citations are all one; however, the distances between their corresponding percentile rank scores are 15.38, 7.7, and 1.92, respectively (see Table 1 and Figure 3). In this case, the sum of the citation count of paper A with 42 raw citations and paper B with 1 raw citation is equal to the sum of the citation count of paper C with 43 raw citations and paper D with 0 citations; however, the sum of the corresponding percentile rank scores of the above papers is no longer equal (see Table 1, equations (3) and (4)). Moreover, as another example, what is even more exaggerated is that the sum of the percentile rank scores of 4 papers with 1 raw citation is greater than the percentile rank score of one paper with 200 citations, which is very unreasonable in actual scientific research evaluation (see Table 1, equations (5) and (6)). Therefore, the nonlinear normalized citation counts of individual papers cannot be added or averaged.

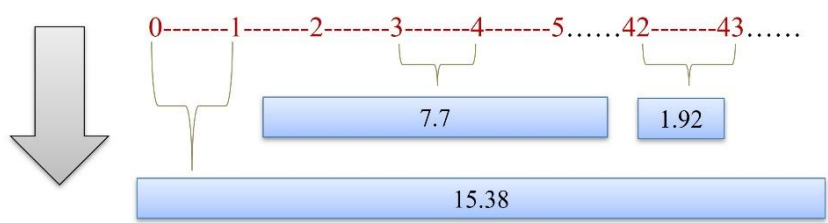

**Figure 3** Schematic diagram showing that each measuring unit of normalized citation counts is no longer equidistant after performing the nonlinear normalization transformation

Before performing the nonlinear transformation:

$$\underset{(A)}{42} + \underset{(B)}{1} = \underset{(C)}{43} + \underset{(D)}{0}, \tag{3}$$

After performing the nonlinear transformation:

$$\underset{(A)}{92.31} + \underset{(B)}{32.69} \neq \underset{(C)}{94.23} + \underset{(D)}{17.31}. \tag{4}$$

Before performing the nonlinear transformation:

$$\underset{(E)}{1} + \underset{(F)}{1} + \underset{(G)}{1} + \underset{(H)}{1} < \underset{(I)}{200}, \tag{5}$$

After performing the nonlinear transformation:

$$\underset{(E)}{32.69} + \underset{(F)}{32.69} + \underset{(G)}{32.69} + \underset{(H)}{32.69} > \underset{(I)}{100}. \tag{6}$$

In fact, the nonlinear field normalization transformation and the nonequidistant transformation are mutually sufficient and necessary conditions (we have proved this theorem in Appendix B). On the one hand, if the mapping relationship between the raw citation counts and their normalized citation counts is nonlinear, then there must exist at least two points that have different slope values in the mapping relationship diagram: as shown in Figure 4, the slope values $k1$ and $k2$ are different. The slope value here is actually the scaling term $k$ mentioned above, and the difference between these scaling terms means that the scaling is not equidistant; that is, the nonlinear field normalization transformation is not an equidistant transformation. On the other hand, if a field normalization transformation is not an equidistant transformation, then there must exist different scaling terms $k$; that is, there must exist different slope values, which will result in a nonlinear mapping relationship between the raw citation counts and their normalized citation counts, as shown in Figure 4.

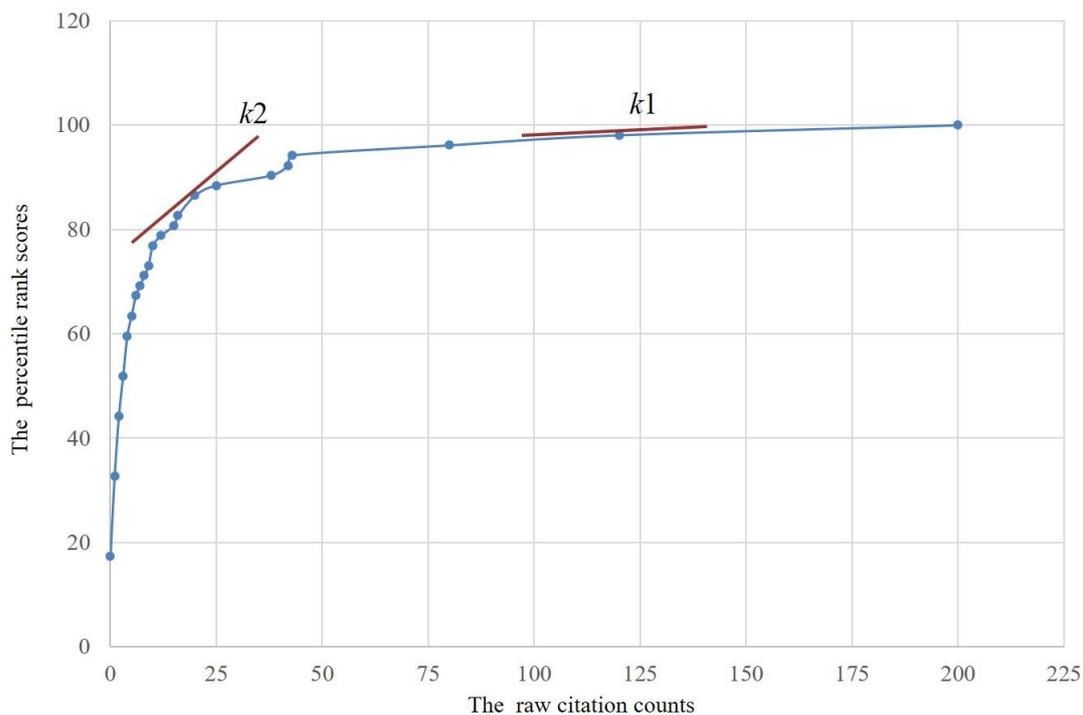

**Figure 4** The mapping relationship between the raw citation counts and the percentile rank scores of 52 papers

## Comb and summary of existing popular linear and nonlinear field normalization methods

In the second section, we have already defined a linear field normalization method and a nonlinear field normalization method. Here, we comb and summarize the existing popular linear and nonlinear field normalization methods.

The mean-based method, median-based method, z score method, optimization-based linear method (Zhang et al., 2015), and Relative Citation Ratio (RCR) method (Hutchins et al., 2016) are all linear field normalization methods. Under these field normalization methods, the mapping from the raw citation count $x$ to the normalized citation count $y$ in the same field follows a linear relationship, such as $y = kx + b$. We summarize the parameters $k$ and $b$ in these linear field normalization methods (see Table 2).

**Table 2** Summary of some major linear field normalization methods

| Name of method | Mathematical expression and brief description of method | k, b |
|---|---|---|
| Mean-based method[4] | $$y = \frac{x}{m} = \frac{1}{m} \cdot x$$ $x$ and $y$ are the raw and normalized citation counts of papers, respectively, and $m$ is the average citation count of papers with the same field as the given paper (the same below). | $k = \frac{1}{m}, b = 0$ |
| Median-based method | $$y = \frac{x}{median} = \frac{1}{median} \cdot x$$ *median* is the median citation count of papers with the same field as the given paper. | $k = \frac{1}{median}, b = 0$ |
| Z score method | $$y = \frac{x-m}{sd} = \frac{1}{sd} \cdot x - \frac{m}{sd}$$ *sd* is the stand deviation of citation counts of papers with the same field as the given paper. | $k = \frac{1}{sd}, b = -\frac{m}{sd}$ |
| Optimization-based linear method | This method first assumes that the citation distribution curves of all fields coincide with a common reference distribution curve after the transformation by the linear method $y = kx + b$ with unknown parameters $k$ and $b$ and then calculates the $k$, $b$ values of each field at this time (this process is equivalent to an optimization problem seeking $k$ and $b$ values when the normalization effect is optimal). When the parameters $k$ and $b$ of each field are obtained, the normalization method $y = kx + b$ of each field is naturally determined. | See the description in the left column |
| Relative Citation Ratio (RCR) method | $$y = \frac{x}{ECR} = \frac{1}{ECR} \cdot x$$ 1. Under this method, the field of one paper is determined by the cocitation network to which this paper belongs. 2. The *FCR* value (the expected field citation count) of one paper is equal to the average of the journal citation rates of all the papers in this focal paper's cocitation network. 3. When using this method, a group of benchmark papers must be selected first and the *ACR* value (the mean number of citations per year) and *FCR* value of each benchmark paper should be calculated. Then, the regression curve (equation) is fitted through the *ACR* values (dependent variable) and *FCR* values (independent variable) of these benchmark papers: $ACR = \beta \cdot FCR + \alpha$  4. In actual evaluation, the *ECR* value (expected citation count) of one focal paper is equal to the *ACR* value when the independent variable of the above regression equation takes the *FCR* value of this focal paper. The final normalized citation count of the focal paper is equal to the raw citation count of this focal paper divided by its expected citation count—*ECR* value. | $k = \frac{1}{ECR}, b = 0$ |

Notably, the RCR method, which has been popular in biomedical circles in recent years, can be regarded as a special linear field normalization method. Under this method, the field to which one paper belongs is determined by this paper's cocitation network; thus, it is difficult for two papers to have exactly the same cocitation network in reality. In other words, there are usually not two or more papers that can belong to the same field in reality: each paper has its own exclusive field. Therefore, under this method, each paper has its own exclusive expected citation count (*ECR* value) and corresponding exclusive scaling term $k$ ($k=1/ECR$).

The percentile rank method, citation z score method (Lundberg, 2007), Normalized Log Citation Score (NLCS) method (Thelwall, 2017), and reverse engineering method (Radicchi & Castellano, 2012a) are all typical nonlinear field normalization methods (see Table 3). Under these field normalization methods, the mapping from the raw citation count $x$ to the normalized citation count $y$ in the same field cannot form a linear relationship such as $y = kx + b$.

**Table 3** Summary of some major nonlinear field normalization methods

| Name of method | Mathematical expression and brief description of method |
|---|---|
| Percentile rank method | The percentile rank score $y$ of one paper's raw citation count $x$ is defined as there are citation counts of $y$ % of the papers with the same field as the given paper are (at or) below the raw citation count $x$ of the given paper. |
| Citation z score method | $$y = \frac{\ln(x+1) - \mu_{[\ln]}}{\sigma_{[\ln]}}$$ $x$ and $y$ are the raw and normalized citation counts of papers, respectively (the same below), $\mu_{\ln}$ and $\sigma_{\ln}$ are the mean and standard deviation of the log-transformed citation counts (plus one) distribution, respectively (the same below). This method is essentially a logarithmic z score method. |
| Normalized Log Citation Score (NLCS) method | $$y = \frac{\ln(x+1)}{\mu_{[\ln]}}$$ This method is similar to the above citation z score method and is essentially a logarithmic mean-based method. |
| Reverse engineering method | $$x = \lambda(y)^{\alpha}$$ The papers located at the same distribution position in different field citation distributions should have the same citation impact. Based on this idea, this method first constructs a common reference distribution using all papers' citations from different fields. When mapping the raw citation count $x$ in a field to the normalized citation counts $y$, the position $P$ of $x$ in the raw citation distribution of this field is first calculated; then, the citation count $y$ located at the same position $P$ in the common reference distribution is found: $y$ is the normalized citation count of $x$. This mapping process above in most fields can be approximately described by the nonlinear power-law function as above, where $\lambda$ and $\alpha$ are field-dependent constants. This method is designed from the perspective of actively meeting the ideal goal of normalization (i.e., all normalized citation distributions of different fields coincide with the same distribution). |

For the percentile rank method, the percentile rank itself is a rectangular distribution by definition, while the distribution of raw citation counts is always a nonrectangular distribution. The mapping from the raw citation counts to the percentile rank scores changes the shape of the nonrectangular distribution of the raw citation counts, so the mapping process is not an equidistant transformation. According to the second section of this paper, the nonequidistant transformation and the nonlinear normalization are mutually sufficient and necessary conditions; therefore, the percentile rank method is a nonlinear field normalization method.

The citation z score method and NLCS method are all logarithmic transformations, as shown in Table 3. The logarithmic transformation is a nonequidistant transformation, and the image of the logarithmic function indicates that logarithmic transformation is obviously a nonlinear transformation. Therefore, the citation z scrore and NLCS methods are all nonlinear field normalization methods.

For the reverse engineering method, as seen in Table 3, the mapping relationship between the raw citation count $x$ and the normalized citation count $y$ in most fields can be described by a nonlinear power-law function; hence, the reverse engineering method is obviously a nonlinear field normalization method.

Moreover, there are three field normalization methods for which it is difficult to determine whether they are linear or nonlinear, namely, the citing-side normalization method (Leydesdorff & Opthof, 2010; Waltman & van Eck, 2013; Zitt & Small, 2008), the method combining citing-side normalization and citation percentile (Bornmann, 2020), and the Exchange Rate method (Crespo et al., 2013) (see Table 4).

**Table 4** Summary of other field normalization methods that are difficult to classify as linear or nonlinear normalization methods

| **Name of method** | **Mathematical expression and brief description of method** |
| --- | --- |
| Citing-side normalization method | The main idea of this method is that each citation is normalized by the number of references in the citing publication or the citing journal, and one significant advantage of this method is that it does not require a field classification system. The three specific approaches of this method are as follows, and SNCS is the abbreviation of "source normalized citation score". $$SNCS1 = \sum_{i=1}^{c} \frac{1}{a_i}, \quad SNCS2 = \sum_{i=1}^{c} \frac{1}{r_i}, \quad SNCS3 = \sum_{i=1}^{c} \frac{1}{p_i r_i},$$ where $c$ is the citation count of the focal paper (the same below), $a_i$ is the mean number of active references in publications which occurred in the same journal and the same publication year as the $i$th citing publication, $r_i$ is the number of active references in the $i$th citing publication, and $p_i$ is the share of publications having at least one active reference among the publications that occurred in the same journal and the same publication year as the $i$th citing publication. |
| The method combining citing-side normalization and citation percentile | This method first normalizes the raw citation counts through the citing-side method, such as SNCS3, and then gives each paper a corresponding percentile rank score according to the SNCS3 score of each paper. |

**Table 4** (Continued)

| Name of method | Mathematical expression and brief description of method |
|---|---|
| Exchange Rate method | The idea of this normalization method is similar to the idea of converting a certain country's currency into a common international currency according to the exchange rate of currency, and the Exchange Rate in this method is similar to the exchange rate between a country's currency and the common international currency.<br><br>**The original Exchange Rate and corresponding field normalization indicator are as follows:**<br><br>$$e_f(\pi) = \frac{\mu_f^{\pi}}{\mu^{\pi}}, \quad y = \frac{x}{e_f(\pi)} \quad (k = \frac{1}{e_f(\pi)}, \ b = 0).$$<br><br>For each field $f$, this method first partitions the papers in field $f$ into $n$ quantile intervals of size $N_f / n$ according to the citation counts of these papers ($N_f$ represents the total number of papers in field $f$; $\pi$ denotes a certain quantile interval and $\pi = 1, 2, \cdots, n$; there are $N_f / n$ papers in each quantile interval $\pi$ in field $f$). Similarly, all papers of all fields are partitioned into $n$ quantile intervals of size $N / n$ according to the citation counts of these papers. $\mu_f^{\pi}$ represents the average citation count of papers in quantile interval $\pi$ of field $f$. $\mu^{\pi}$ represents the average citation count of papers in quantile interval $\pi$ of the citation distribution composed of all papers in all fields, $e_f(\pi)$ is the Exchange Rate for the papers in the quantile interval $\pi$ of field $f$ (the same below). $x$ is the raw citation count of one focal paper that belongs to the quantile interval $\pi$ of field $f$, and $y$ is the normalized citation count of this focal paper.<br><br>**The redefined Exchange Rate and corresponding field normalization indicator are as follows:**<br><br>$$e_f = \frac{1}{(\pi^M - \pi_m)} \sum_{\pi} e_f(\pi), \quad y = \frac{x}{e_f} \quad (k = \frac{1}{e_f}, \ b = 0).$$<br><br>$\pi^M$ ($\pi_m$) is the upper (lower) boundary value of a special interval $[\pi_m, \pi^M]$, and $\pi \in [\pi_m, \pi^M]$. $e_f$, which is equal to the arithmetic mean of the original Exchange Rate for every quantile interval $\pi$ in this special interval, is the Exchange Rate for all papers belonging to field $f$. $x$ is the raw citation count of one focal paper belonging to field $f$, and $y$ is the normalized citation count of this focal paper. |

The concept of linear and nonlinear field normalization methods is premised on a field classification system. Before performing the normalization transformation, the raw citation counts within the same field are comparable and additive. After performing the linear field normalization transformation, the normalized citation counts within the same field are still comparable and additive (the normalized citation counts from different fields are also certainly comparable and additive at this time because the field normalization transformation has been conducted). After performing the nonlinear field normalization transformation on the raw citation counts, the normalized citation counts within the same field cannot be added anymore according to the analysis of the above second section. In terms of the citing-side normalization method, this method itself has no concept of field classification, and the normalization process does not require a field classification system. Therefore, this method itself falls outside the conceptual category of distinguishing linear and nonlinear field normalization methods; it is difficult for us to classify this method as a linear or nonlinear field normalization method.

The method combining citing-side normalization and citation percentile is essentially a citing-side normalization method. However, for the convenience of people to better understand the normalized citation counts under the citing-side method such as SNCS3, this method finally gives each paper a corresponding percentile rank score according to the SNCS3 score of each paper. Since this method is essentially a citing-side normalization method, this method also does not have the concept of the field classification system, and its normalization process does not need a field classification system. Thus, this method also falls outside the conceptual category of distinguishing linear and nonlinear field normalization methods.

The Exchange Rate method is a nonlinear field normalization method from the perspective of the original definition of this method, and the reasons can be explained by analyzing the formula of the original Exchange Rate $e_f(\pi) = \dfrac{\mu_f^{\pi}}{\mu^{\pi}}$ (see Table 4). The average citation counts of papers in different quantile intervals $\pi$ within the same field (i.e., the numerator of the Exchange Rate) are different, for example, $\mu_f^1 \neq \mu_f^{200}$. Similarly, the average citation counts of papers in different quantile intervals $\pi$ of the citation distribution composed of all papers in all fields (i.e., the denominator of the Exchange Rate) are different, for example, $\mu^1 \neq \mu^{200}$. Accordingly, the Exchange Rates at different quantile intervals $\pi$ within the same field will not be exactly the same with high probability, for example, $\dfrac{\mu_f^1}{\mu^1} \neq \dfrac{\mu_f^{200}}{\mu^{200}}$, i.e., $e_f(1) \neq e_f(200)$. That is, the scaling term $k = \dfrac{1}{e_f(\pi)}$ of the raw citation counts of papers at different quantile intervals $\pi$ within the same field are different (e.g., $\dfrac{1}{e_f(1)} \neq \dfrac{1}{e_f(200)}$), which indicates that this scaling is not equidistant; thus, this method is a nonlinear field normalization method.

However, possibly due to the convenience of calculation and other reasons, Crespo et al. selected a certain special interval $[\pi_m, \pi^M]$ such as $[706, 998]$ to redefine the Exchange Rate: $e_f = \dfrac{1}{(\pi^M - \pi_m)} \sum_\pi e_f(\pi)$ (see Table 4), which is equal to the arithmetic mean of the original Exchange Rate for every quantile interval $\pi$ in this special interval $[\pi_m, \pi^M]$ and is used as the uniform Exchange Rate for all papers belonging to field f. In this case, the scaling term $k$ of the raw citation counts of all papers within the same field is a uniform value equal to $\dfrac{1}{e_f}$; thus, this scaling is equidistant, and this transformation process is naturally linear.

Therefore, in summary, for the Exchange Rate method, whether the method is a linear or nonlinear field normalization method must be determined according to the specific algorithm definition.

## Conclusions and discussions

In this paper, we performed the following two research works, which are also two academic contributions of this paper. First, from the perspective of theoretical analysis in mathematics, we analyzed why the nonlinear normalized citation counts of individual papers obtained using nonlinear field normalization methods cannot be added or averaged, which has not been done before in academia: the prerequisite that the data can be added is that the measuring unit of the data is equidistant; however, the normalization procedure of the nonlinear field normalization method is not an equidistant transformation (for this crucial step that the nonlinear field normalization method is not an equidistant transformation, we provide a mathematical proof); thus, the measuring unit of nonlinear normalized citation counts is no longer equidistant. Therefore, the nonlinear normalized citation counts of individual papers cannot be added or averaged. Second, we systematically classified the existing main field normalization methods into linear and nonlinear field normalization methods, which also has not been done before in academia.

The above two research works provide a further theoretical basis for the proper use of field normalization methods in research evaluation in the future, avoiding the continued misuse of nonlinear field normalization methods in academic and evaluation practice circles.

Furthermore, the proofs of Theorem 1 and Theorem 2 in our appendices aim to address not only the specific topic of the relationship between linear (nonlinear) field normalization transformation and the equidistant (nonequidistant) transformation but also the relationship between linear (nonlinear) transformation and the equidistant (nonequidistant) transformation over the entire real number domain in a more general sense. Therefore, we actually proved that all nonlinear data in any real number domain (such as percentile rank scores in other research topics) cannot be added or averaged, which is meaningful for the whole field of data science and information science.

In fact, not only nonlinear field normalization methods have long been misused, but also other nonlinear data, such as percentile rank scores and logarithmic scores in other research topics belonging to the field of data science and information science, have always been misused because these nonlinear data have also been added or averaged. The misuse of nonlinear data appears frequently even in some of the top journals of information science (e.g., Jackson et al., 2019; Safon & Docampo, 2020; Zheng et al., 2022). Therefore, our research work in this paper is meaningful not only to the theory and practice of the field of field normalization, research evaluation, science of science, and informetrics but also to the whole field of data science and information science.

# Endnotes

[1] The comparison of citation impact across fields and the corresponding question of field normalization are often necessary and hard to avoid in actual research evaluation practice because the actual research evaluation is often carried out at aggregation levels (e.g., at the level of institutions, journals and countries). These evaluated objects, such as institutions, are rarely monodisciplinary and often have many research fields (Bornmann, 2020; Radicchi & Castellano, 2012b; Rafols et al., 2012). Furthermore, competitions for the same resources from scholars in different fields, such as academic positions or academic awards, are also involved in the conundrum of comparing apples with oranges (Radicchi & Castellano, 2012b).

[2] There are various specific methods for calculating the percentage rank score, such as the $P100$, $P100$', $CP-IN$ and $CP-EX$ methods. For these specific methods, please refer to the paper entitled "An evaluation of percentile measures of citation impact, and a proposal for making them better" (Bornmann & Williams, 2020).

[3] The specific method for calculating the percentile rank scores here is the $CP-IN$ method (Bornmann & Williams, 2020).

[4] Here, we only comb the field normalization methods based on the paper level in this section and this table, so some mean-based normalization indicators at the aggregation level, such as the new crown indicator (Waltman et al., 2011) and WCNCI indicator (Wang & Zhang, 2020), are not listed separately. Similarly, other field normalization indicators at the aggregation level, such as the $I3$

indicator (Leydesdorff & Bornmann, 2011), are not listed separately in this section and the following tables.

# Appendix A: The proof of Theorem 1

In this Appendix, we provide a proof of Theorem 1.

**Theorem 1.** The linear transformation and the equidistant transformation are mutually sufficient and necessary conditions.

**Proof** We divide our proof into two steps. First, we need to prove that the linear transformation must be an equidistant transformation, that is, the linear transformation $\Rightarrow$ the equidistant transformation.

Let $y = f(x) = kx + b$ denote an arbitrary linear transformation method, where $x$ is the value before performing the linear transformation and its definitional domain is the real number domain (in the context of this paper, it refers to the raw citation count of one paper), $y$ is the value after performing the linear transformation (in the context of this paper, it refers to the normalized citation count of this paper), $k$ is a nonzero constant and $b$ is a constant.

Then, let us choose any four values $x_i$, $x_j$, $x_m$ and $x_n$ on the real number domain of the independent variable $x$, such that

$$x_j - x_i = x_n - x_m \ (x_j > x_i, x_n > x_m). \tag{7}$$

That is, we suppose that the values of the independent variables are equidistant (in the context of this paper, it means that each citation count of the raw citation count is equidistant before performing the linear normalization transformation).

Next, our task is to compute the difference between $y_j$ and $y_i$, and we obtain

$$\begin{aligned} y_j - y_i &= f(x_j) - f(x_i) \\ &= kx_j + b - (kx_i + b) \\ &= k(x_j - x_i). \end{aligned} \tag{8}$$

By substituting equation (7) into equation (8), we have

$$\begin{aligned} y_j - y_i &= k(x_j - x_i) \\ &= k(x_n - x_m) \\ &= kx_n - kx_m \\ &= (kx_n + b) - (kx_m + b) \\ &= y_n - y_m \end{aligned} \tag{9}$$

The equation $y_j - y_i = y_n - y_m$ implies that the values of the variables are still equidistant after performing the linear transformation. We have thus proved that the linear transformation must be an equidistant transformation (similarly, in the context of this paper, let the above $x_i, x_j, x_m$ and $x_n$ be raw citation counts such that $x_i, x_j$ are adjacent items and $x_m, x_n$ are adjacent items, i.e., $x_j - x_i = x_n - x_m = 1$ $(x_j > x_i, x_n > x_m)$ ; then, substituting this equation into the above equation (8), we have $y_j - y_i = y_n - y_m = k$ , which implies that each measuring unit of normalized citation count is still equidistant after performing the linear normalization transformation and the distance equals $k$).

Second, the next step in the proof is to prove that the equidistant transformation must be a linear transformation, that is, the equidistant transformation $\Rightarrow$ the linear transformation.

Let $y = f(x)$ denote an arbitrary equidistant transformation method, where $x$ is the value before performing the equidistant transformation and its definitional domain is the real number domain (in the context of this paper, it refers to the raw citation count of one paper), $y$ is the value after performing the equidistant transformation (in the context of this paper, it refers to the normalized citation count of this paper) and $f$ is an equidistant mapping.

Next, as shown in Figure 5, we choose any three adjacent values $x_i, x_j$ and $x_m$ on the real number domain of the independent variable $x$, such that

$$x_m - x_j = x_j - x_i \quad (x_m > x_j > x_i) . \tag{10}$$

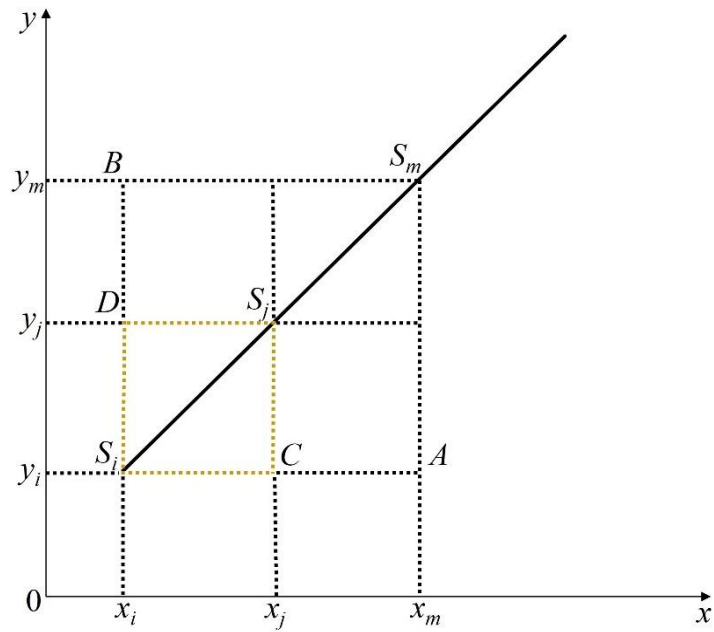

**Figure 5** The first schematic diagram for helping to prove that "the equidistant transformation must be a linear transformation"

That is, we first suppose that the adjacent values of the independent variables are equidistant (in the context of this paper, it means that each citation count of the raw citation count is equidistant before performing the normalization transformation). Since $f(x)$ is an equidistant transformation, it follows that

$$y_m - y_j = y_j - y_i \tag{11}$$

(in the context of this paper, it means that each measuring unit of the normalized citation count is still equidistant after performing the normalization transformation; $S_j$ is the center of the rectangle $S_i A S_m B$ ) and it must be a line segment between endpoints $S_i$ and $S_j$, and it must be a line segment between endpoints $S_j$ and $S_m$ (see Figure 5).

If the assertion that the distance between endpoints $S_i$ and $S_j$ must be a line segment would not hold, and if we also want to ensure that the function $f(x)$ is still an equidistant transformation at this time, then the image of the function $f(x)$ must pass through the center of the rectangle $S_i C S_j D$, which thus leads to $x_j - x_1 = x_1 - x_i$ and $y_j - y_1 = y_1 - y_i$. The distance between endpoints $S_i$ and $S_1$ and the distance between endpoints $S_1$ and $S_j$ can be a curve of any shape (see Figure 6).

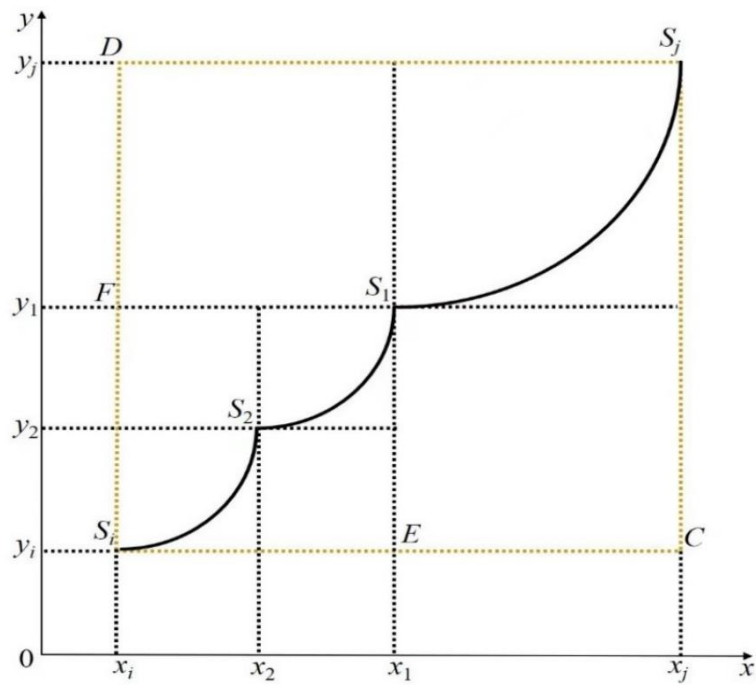

**Figure 6** The second schematic diagram for helping to prove that "the equidistant transformation must be a linear transformation"

Similarly, if we want to ensure that the distance between endpoints $S_i$ and $S_1$ is not a line segment and want to ensure that the function $f(x)$ is still an equidistant transformation at this time, then the image of the function $f(x)$ must pass through the center of the rectangle $S_iES_1F$, which thus leads to $x_1 - x_2 = x_2 - x_i$ and $y_1 - y_2 = y_2 - y_i$. The distance between endpoints $S_i$ and $S_2$ and the distance between endpoints $S_2$ and $S_1$ can be a curve of any shape (see Figure 6).

Following the same procedure as above, there must exist a value $x_{n+1}$ such that $x_n - x_{n+1} = x_{n+1} - x_i$, which leads to $y_n - y_{n+1} \neq y_{n+1} - y_i$, i.e., the image of the function $f(x)$ does not pass through the center $S'_{n+1}$ of the rectangle $S_iGS_nH$ (see Figure 7). This contradicts the fact that the function $f(x)$ is an equidistant transformation. Therefore, the distance between endpoints $S_i$ and $S_n$ must be a line segment. Similarly, it can be proven that the distance between endpoints $S_n$ and $S_{n-1}$ must be a line segment. Since $S_n$ is the center of the rectangle $S_iIS_{n-1}J$, it follows that endpoints $S_i$, $S_n$ and $S_{n-1}$ are all on a line segment.

Based on the same logic, we can use step-by-step backward deduction to show that the distance between endpoints $S_i$ and $S_m$ must be a line segment and endpoints $S_i$, $S_j$ and $S_m$ are all on this line segment.

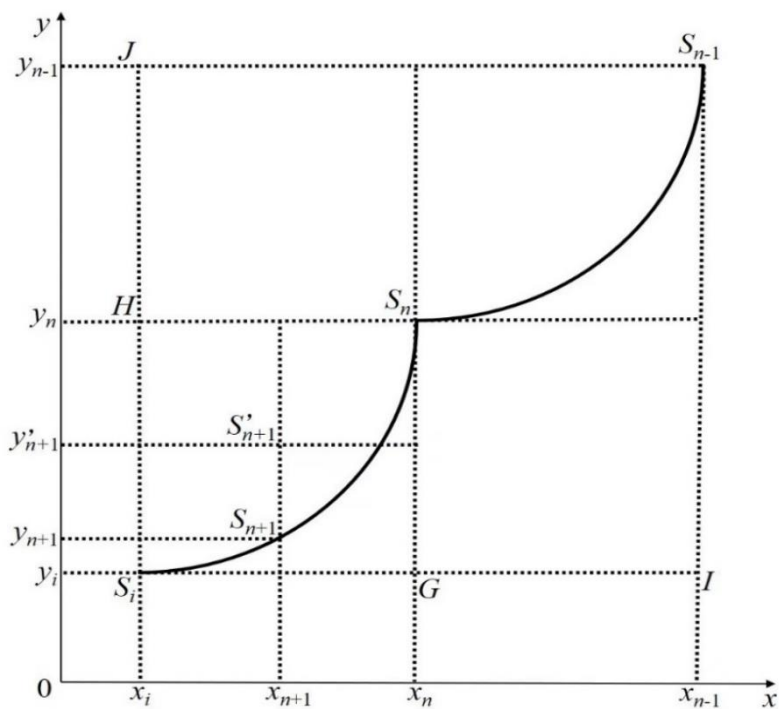

**Figure 7** The third schematic diagram for helping to prove that "the equidistant transformation must be a linear transformation"

Since numerous adjacent points, such as $x_i$, $x_j$ and $x_m$, exist on the real number domain of the independent variable $x$, it can be deduced by analogy from the above proof process that numerous line segments exist, such as $S_i\,S_j$, $S_j\,S_m$, and $S_i\,S_m$; that is, line segment $S_i\,S_m$ can be extended indefinitely and thus is a straight line. We have thus proved that the equidistant transformation $f(x)$ must be a linear transformation.

The proof of Theorem 1 is now complete.

## Appendix B: The proof of Theorem 2

In this Appendix, we provide a proof of Theorem 2.

**Theorem 2.** The nonlinear transformation and the nonequidistant transformation are mutually sufficient and necessary conditions.

**Proof** Since we have proved that the linear transformation must be an equidistant transformation (i.e., the linear transformation $\Rightarrow$ the equidistant transformation) in Appendix A, the contrapositive of the above conclusion is also true, that is, the nonequidistant transformation must be a nonlinear transformation (i.e., the nonequidistant transformation $\Rightarrow$ the nonlinear transformation). In the context of this paper, the nonequidistant transformation must be a nonlinear field normalization transformation.

Similarly, because we have also proved that the equidistant transformation must be a linear transformation (i.e., the equidistant transformation $\Rightarrow$ the linear transformation) in Appendix A, the contrapositive of the above conclusion is also true, that is, the nonlinear transformation must be a nonequidistant transformation (i.e., the nonlinear transformation $\Rightarrow$ the nonequidistant transformation). In the context of this paper, the nonlinear field normalization transformation must be a nonequidistant transformation.

Therefore, the nonlinear transformation and the nonequidistant transformation are mutually sufficient and necessary conditions. The proof of Theorem 2 is complete.

# Supporting Information

Xing Wang[a,1]

[a]School of Information, Shanxi University of Finance & Economics, 030006 Taiyuan, China

### *A discussion about the dilemma regarding the use of linear and nonlinear field normalization methods*

According to the analysis of the second section of this paper and the proofs of Theorems 1 and 2 in the Appendices, we know that we should select and use the linear field normalization method in research evaluation activities at aggregation levels. However, there is still a dilemma regarding the choice and use of linear and nonlinear field normalization methods that needs to be discussed, starting with discussing of the test method of the normalization effect and the ideal goal of field normalization.

The crucial criterion to judge the performance of one field normalization method is the normalization effect of the method, i.e., whether the method can effectively suppress the field bias of raw citation counts and thus make the normalized citation counts independent of field. "Effectively suppress the field bias of raw citation counts and thus make the normalized citation counts independent of field allowing for a fair comparison of research impact of papers from different fields" is the ideal goal of field normalization. The following three test methods are summarized by Bornmann (2018) to examine the normalization effect of field normalization methods, namely, the fairness test method (Radicchi et al., 2008), the intraclass correlation coefficient (ICC) test method (Leydesdorff & Bornmann, 2011), and the inequality due to differences in citation practices (IDCP) term test method (Crespo et al., 2013; Li & Ruiz-Castillo, 2013). Among these test methods, the fairness test method is the most commonly used and easily understood, so this paper focuses on only this method.

The fairness test method is divided into the following two specific approaches. The first approach is named the top z% approach, which examines the normalization effect of the normalization method from a numerical quantitative perspective. This approach first ranks all papers from all fields in descending order according to the papers' normalized citation counts and then extracts the top z% of all papers. Next, we compute the proportion of papers in each field that fall into the top z%. If the normalization method is fair and effective, the proportion for all fields should be z% because the above proportion should not depend on the specific field to which the paper belongs but only relate to the number of papers in the field to which the paper belongs. The closer the proportion of each field is to z%, the more effective the normalization method is. The second approach is named the probability distribution plots approach, which examines the normalization effect of the normalization method from a graphical perspective.

Under this test approach, if the normalization method is fair and effective, the normalization method should be able to make all normalized citation distributions of different fields coincide with the same distribution. The closer the normalized citation distribution curves of different fields are to each other, the more effective the normalization method is.

Clearly, from the perspective of achieving the ideal goal of field normalization (i.e., effectively suppressing the field bias of raw citation counts, thus making the normalized citation counts independent of the field and allowing for a fair comparison of the research impact of papers from different fields), the percentile rank method is an ideal normalization method because its definition itself conforms to the idea of the fairness test method—top z% approach. That is, the percentile rank method is a normalization method designed from the final normalization effect, i.e., the final ideal goal of normalization. Similarly, the reverse engineering method is also an ideal normalization method because its definition itself also conforms to the idea of the fairness test method—the probability distribution plots approach.

This situation will bring a dilemma regarding the use of field normalization methods. On the one hand, the percentile rank method and the reverse engineering method both have the ideal normalization effect because they themselves are designed from the final ideal goal of normalization, especially the percentile rank method, which also has the advantages of being simple and easy to understand. On the other hand, however, the normalized citation counts at the paper level obtained using the percentile rank method or the reverse engineering method, which are nonlinear normalization methods, cannot be added or averaged. Therefore, the percentile rank method and the reverse engineering method naturally cannot be used for research evaluation at aggregation levels where is the most commonly used scenario at actual research evaluation activities.

One possible compromise to solve this dilemma is to use the optimization-based linear method proposed by Zhang et al. (2015). This method, enlightened by the idea of the reverse engineering method, is also designed from the final ideal goal of field normalization, so this method also has a good normalization effect. Furthermore, this method is a linear field normalization method, so it can naturally be used in research evaluation at aggregation levels.

However, the greatest shortcoming of the optimization-based linear method is that its algorithmic process is too complicated and difficult to apply in practical research evaluation activities. Moreover, because the shape changes of the normalized citation distribution curves via a linear transformation are not as flexible as those via a nonlinear transformation, the normalized citation distribution curves via the optimization-based linear method cannot completely coincide with the reference distribution curve (Zhang et al., 2015), so the optimization-based linear method does not show decided advantages over the mean-based method, which is another main shortcoming of this linear normalization method. Nevertheless, the slight advantage of the optimization-based linear method designed from the final ideal goal of field normalization over the mean-based method exactly implies that the mean-based method also has a good normalization effect.

Some scholars believe that since citation distributions are usually skewed, the median is a more appropriate measure of central tendency than the arithmetic average, that the median-based method is a more reasonable method than the mean-based method, and that the median-based method should be used instead of the mean-based method in field normalization. However, the performance of one field normalization method should be judged by the normalization effect of this method or by investigating whether this method achieves the final ideal goal of field normalization—effectively suppressing the field bias of raw citation counts and thus making the normalized citation counts independent of field, rather than whether this method uses the central tendency of the citation distributions. There is no theoretical basis for whether using better central tendency parameters can result in a better normalization effect. In fact, the median of citation data is usually smaller than the arithmetic mean. If the median-based method is used, those high citation counts located in the head of the citation distribution will be divided by a relatively smaller value than the arithmetic mean, and thus, the field advantage of these high raw citation counts cannot be effectively suppressed.

In summary, comprehensively considering the following factors, including the normalization effect, the algorithmic complexity, and the application scenarios of the field normalization method, such as whether the normalization method can be applied at the aggregation level, the mean-based method, as a linear normalization method, may already be a very good normalization method in practical research evaluation activities at the present stage and may achieve a balance in terms of the dilemma of the use of the field normalization method.

*Scientometrics, 102*(1), 587-607.